# Techno-Economic Assessment in Communications: New Challenges

Carlos Bendicho[1]  and Daniel Bendicho[2]

[1] Independent ICT Researcher, Member, IEEE, Member, ACM, Member, FITCE, COIT, Spain
[2] Independent Researcher, Spain
`carlos.bendicho@coit.es`

**Abstract.** This article shows a brief history of Techno-Economic Assessment (TEA) in Communications, a proposed redefinition of TEA as well as the new challenges derived from a dynamic context with cloud-native virtualized networks, the Helium Network & alike blockchain-based decentralized networks, the new network as a platform (NaaP) paradigm, carbon pricing, network sharing, and web3, metaverse and blockchain technologies. The authors formulate the research question and show the need to improve TEA models to integrate and manage all this increasing complexity. This paper also proposes the characteristics TEA models should have and their current degree of compliance for several use cases: 5G and beyond, software-defined wide area network (SD-WAN), secure access service edge (SASE), secure service edge (SSE), and cloud cybersecurity risk assessment. The authors also present TEA extensibility to request for proposals (RFP) processes and other industries, to conclude that there is an urgent need for agile and effective TEA in Comms that allows industrialization of agile decision-making for all market stakeholders to choose the optimal solution for any technology, scenario and use case.

## 1 Introduction

The advent of new communications, computing and cyber security technologies such as 5G, 6G, software-defined networking (SDN), network function virtualization (NFV), secure access service edge (SASE) and secure service edge (SSE) has increased the technical complexity and diversity of architectural solutions.

Thus, decision makers need effective Techno-Economic Assessment (TEA) models to choose the optimal solution.

Current techno-economic models in communications prioritize economic feasibility instead of both technical and economic viability of complex technical solutions.

Besides, they are mainly focused on network deployment and telecom operators' perspectives and seldom consider end customers´ needs.

Therefore, Assessment in Comms is becoming a crucial need for organizations and it should be considered in the same way that assessment in cloud adoption and migration is included in the current portfolios of service providers and consulting companies.

For instance, techno-economic assessment will allow to address, in optimal conditions, the transformation of a more than $100 billion global market of managed network services by 2030 at 6% CAGR from 2021 [1], considering different flavors for SD-WAN, SASE or SSE adoption. And this is only an example of one domain in communications. We can extend it to any other domain as, for example, 5G infrastructure and services market with a $1 trillion investment forecast by operators from 2019 to 2025 as stated by GSMA [2].

Current techno-economic analysis models and tools are especially focused on network deployment considering telecom operators´ perspective [3]. However, a techno-economic assessment can also be applied to the deployment or adoption of any other new access network technology, backhaul, transport or any complete technical solution, such as aerial base stations deployment for emergencies or events coverage, for any market stakeholder: vendors, telecom operators, communications service providers (CSPs), mobile network operators (MNOs), mobile virtual network operators (MVNOs), multinational corporations (MNCs), enterprise organizations, small and medium enterprises (SMEs), public administration, technological consulting companies, regulators, and so on.

Hence, there is a need for solvent and agile techno-economic assessment (TEA) models and tools that allow industrialization of agile decision-making for all market stakeholders to select the optimal solution for any technology, scenario and use case.

This paper is structured as follows. Section 2 presents a brief history of TEA in Communications summarizing the review of the literature. Section 3 shows a proposal of redefinition of TEA. Section 4 exposes new challenges for TEA derived from the evolving technological and market context. In Section 5 the authors formulate the research question. Section 6 shows some use cases of TEA application (5G and beyond, SD-WAN, SASE and SSE, Cybersecurity Risk Assessment), the characteristics TEA models should have and related gaps detected in the literature. Section 7 exposes some current challenges concerning RFP processes that could be addressed by using TEA. Section 8 introduces the extension of TEA models for Comms to other domains and industries, and eventually Section 9 shows the conclusions.

## 2       Brief History of TEA in Comms

Summarizing the review of the literature presented in detail by the author in [3] and [4], there was an American U.S. seed in the field of techno-economic modeling for access networks in the late 1980s and the early 1990s. In 1989, a publication based on dynamic programming predicted that the most appropriate moment to invest massively in FTTH deployment would not be before 2010, considering the forecasting of costs, income and interest rates [5]. In the 1990s, studies began to focus on the detailed cost analysis of components with a "bottom-up" approach always oriented to the deployment of access networks and ignoring the end-user perspective.

In the 1990s, techno-economic modeling for access networks also germinated in Europe with European Union (EU) public-funded projects aimed at choosing the most appropriate alternative for access network deployment by telecom operators and promoting standards and recommendations, with the surge of STEM, TITAN, OPTIMUM, TONIC, and ECOSYS models. The ECOSYS model resulted in a dissemination effect materialized with the emergence of new proprietary models for power line communications (PLC), optical networks, 3G-LTE, the BATET model that distinguished between fixed and nomadic (mobile) layers, the COSTA model based on MUSE (a TONIC model extension), and more models for hybrid FTTx and WiFi or WiMAX networks, fiber to the distribution point (FTTdp). Since 2008, more EU public-funded projects continued the research: BONE, OASE for optical networks TEA, and DISCUS, provoking another spread effect towards proprietary models for deployment of broadband in rural areas, comparing fiber to the home (FTTH) and hybrid fiber coax (HFC) DOCSIS 3.0, for hybrid fiber & WiFi networks, LTE, converged access networks FTTH/FTTB and LTE, and eventually for 5G deployment [6]. In 2019, reference [6] included open source code in the GitHub public repository. Previous TEA models resulting from publicly funded EU projects are made available to the public by contacting project interlocutors. All models in the literature are based on the traditional definition of TEA [7] and mainly consider the deployment of access network technologies from the perspective of operators, vendors, manufacturers and standardization bodies.

## 3       Redefinition of Techno-Economic Assessment (TEA)

The traditional definition of Techno-Economic Assessment is stated as follows: "Techno-economic models are a method used to evaluate the economic viability of complex technical systems", according to Smura's doctoral thesis [7], which obviates any evaluation of technical viability.

Aligned with this definition, current decision-making processes regarding the choice of networking technologies are based almost exclusively on economic criteria, which bears the risk of committing serious technical errors that may compromise expected economic viability.

TEA models in the literature use economic output parameters like capital expenses (CapEx) and operation expenses (OpEx) or even Net Present Value (NPV) and IRR (Internal Rate of Return) to evaluate economic feasibility of a project, but they rarely use technical input or output parameters to evaluate the technical suitability of the proposed system. Most of TEA models just calculate costs in a given scenario without considering technical requirements and end users´ needs.

Therefore, a broader extension of the concept of techno-economic analysis or assessment (TEA) is required. The new definition must emphasize the evaluation of technical feasibility and must be supported by techno-economic models that develop it.

From this perspective, the author proposed a new definition in [3] as follows: "The techno-economic models are methods that allow the evaluation of the technical and economic viability of complex technical systems." This new definition emphasizes both the technical and economic aspects of modeling, considering the technical feasibility and satisfaction of specific technical and economic requirements and needs.

On the other hand, techno-economic models in the literature are eminently oriented towards the dynamics of the deployment of access networks promoted by vendors, manufacturers and operators, ignoring the perspective of end users.

Hence, there is a need for techno-economic assessment models that reflect and respond to both perspectives to contribute to market equilibrium.

## 4   New challenges for TEA in Comms

The context in communications is extraordinarily dynamic incorporating continuously new methods, technologies and innovations that pose new impacts and implications for TEA. This section shows some of these innovations and their implications for TEA in communications.

### 4.1   Validation, Test and Assurance on new Cloud-Native Networks

A new paradigm for designing and operating networks is derived from validation, testing, and assurance needs to deliver high availability in large, complex cloud-native virtualized networks based on IntOps (Integrator and Operator) best practices supported by GitOps methodology and technology that provide workflows for infrastructure and applications with continuous evolution and continuous integration and delivery (CI/CD) [8].

GitOps is a set of practices to manage infrastructure and applications configurations using Git, which is an open source releases control system, as the only source of truth for declarative infrastructure and applications [8].

IntOps is a shift in strategy in progress of adoption by many telecom operators in order to become both integrators and operators [8].

The engineering and quality assurance (QA) areas must design and integrate dynamically and as much automated as possible new cloud-native network functions (CI: continuous integration), which need to be automatically tested for carrier-degree scalability, security, capacity, performance, quality, availability, resilience and compliance (CT: continuous testing) to be delivered end-to-end along the network (CD: continuous delivery).

Continuous testing (CT) is therefore necessary and critical every time a cloud-native network function changes. For instance, every time there is a change in the containers image registry for a given cloud-native network function that will be executed on Kubernetes (containers´orchestrator) clusters (nodes) or a new cloud-native network function is created. CT is required even one step before the image is uploaded in the registry, and also afterwards during execution time to detect any failure and provide feedback for a new version, considering both the customer experience and runtime environment.

Both CI/CD and CT pose new technical and economic impacts and implications for the techno-economic assessment (TEA) of network and security solutions for any market stakeholder. They require capital expenses (CapEx) and operation expenses (OpEx) for tools and knowledge acquisition as well as cultural change. In the long term, continuous integration, testing and delivery (CI/CT/CD) implies OpEx reduction, although CT impacts CapEx and requires smaller orders of magnitude in OpEx.

### 4.2 Blockchain-based Decentralized Networks: the Helium Network & alike

The Helium Network [9] is a decentralized wireless network initially conceived for Internet of Things (IoT) devices and sensors, using LoRaWAN open-source wireless protocol, based on blockchain with a native protocol, called the Helium Consensus protocol, which serves to feed a two-sided market integrated by coverage providers and coverage consumers.

LoRaWAN specification is a Low Power, Wide Area Network open-source protocol designed to connect battery operated devices wirelessly to the internet in regional, national or global networks [9].

The Helium Network is now extending to provide a 5G Citizen Broadband Radio Service (CBRS) spectrum decentralized network called Helium 5G powered by coverage providers´ hotspots.

Coverage providers, who in Helium are the so-called miners in the blockchain terminology, purchase access node equipment, connect it via the Internet and demonstrate their contribution to Helium wireless network coverage in a cryptographically verified physical location and time by a Proof-of-Coverage algorithm, submitting proofs to the Helium network from their own access node.

In the Helium network, IoT or 5G CBRS devices pay to send and receive data via the Internet by using the Helium distributed access network provided by coverage providers, who earn tokens for providing wireless network coverage for those devices, from data traffic transactions passing through their access nodes and for validating the integrity of the Helium network.

The result is a decentralized wireless access network that is commoditized by competition, and provides global coverage at a fraction of the current costs.

Therefore, blockchain-based decentralized networks such as Helium pose a new paradigm that must be considered in TEA from both technical and economic perspectives, as this type of decentralized networks will impact technical functionalities and performance as well as OpEx.

There are also regulatory aspects and uncertainties to consider for TEA as these blockchain-based networks and markets have not been regulated to date.

### 4.3 Evolution of Networks: Network as a Platform (NaaP)

The advent of fully programmable networks that expose capabilities via application programmable interfaces (API) to developers and applications in cloud hyperscalers' marketplaces will provide an additional source of revenues in addition to business-to-business (B2B) or business-to-consumer (B2C) traditional markets for telecom operators, converting telco networks in platforms in the middle of two-sided markets.

This is coming in an iterative cycle, and this time, the materialization of the Network as a Platform (NaaP) paradigm is much closer as GSMA operators have joined the Linux Foundation in the so-called CAMARA community. CAMARA was launched in MWC2022 for operator platform API development and standardization following the "code-first" principle applying agile methodologies. Telecom operators within the GSMA Operator Platform API Group (OPAG) have been advancing in recent years, and many have internal APIs available for the CAMARA community.

These network as a platform capabilities will facilitate a continuous and dynamic dialogue between business, information technology (IT) and operation technology (OT) applications with the network, by means of which network congestion state will be provided to applications so that they can invoke, when appropriate, better network performance (QoS), network latency, network latency stability services and so on. This will allow communications service providers (CSPs) to monetize such network services from application developers and over-the-top companies (OTTs), adding these revenues to those coming from traditional B2B, B2C and wholesale markets.

On the other hand, network as a platform requires a secured API exposure layer to minimize threats and vulnerabilities and grant secure access via appropriate identity and access management (IAM), directly impacting OpEx.

Therefore, the new network as a platform (NaaP) paradigm impacts revenues and OpEx dimensions of TEA and at the same time, these new network capabilities pose new technically advantageous features to be considered and weighed from the technical perspective in techno-economic assessment.

### 4.4 Sustainability Issues: Carbon Price

The objective of net zero emissions under the United Nations (UN) Sustainable Development Goals (SDG) drives governance, risk and compliance (GRC) objectives, environment, social and governance (ESG) criteria for sustainability and environmental governance in stakeholders´ organizations. One measure adopted in this direction is carbon pricing calculation and its inclusion as OpEx in any project economics.

Therefore, the best practice will be to add carbon pricing as an OpEx in techno-economic assessment.

On the other hand, the inclusion per se of carbon pricing information in a solution to be assessed, constitutes a technical input parameter to be considered and weighed in techno-economic assessment versus other alternative solutions.

### 4.5 Network sharing

Telecom operators are deploying different strategies in order to share network infrastructure either with other network operators or non-operator companies from different sectors (energy companies, media companies, hyperscalers, municipalities, private equity, etc.), mainly seeking cost savings as well as quicker rollouts.

Hence, network sharing has an impact on TEA CapEx and OpEx calculations as well as on the technical viability as it makes operators face new challenges by increasing operational complexity, diminishing services differentiation opportunities or control over the network, increasing also the probability of technical incompatibilities with partner´s equipment.

### 4.6 Metaverse, web3 and blockchain

The development and future evolution of the metaverse will increase pressure on computing power and networks as it needs large-data processing such as graphics rendering for the myriad of virtual worlds that will emerge. The authors envision the metaverse as one of the future 5G and beyond massive use cases gathering consumer and enterprise users around virtual, augmented and extended reality mobile devices.

There are estimates about $5 trillion metaverse value generation by 2030 [16] and even 4 hours of metaverse presence per capita [17].

Massive interconnected metaverses will require real time interactions processing considering navigation, tactile and haptic interactions including persistence and footprint of users in every metaverse. Today, there are 5.4B internet mobile connections per day [GSMA] what means that more than 5B billion users will be able to access thoses metaverses.

Moreover, the development and evolution of web3 and blockchain technologies associated or not with the metaverse will add even more pressure on computing power and networks. There are forecasts about 20x data and 1000x computing power required.

Telcos will need to deploy massive edge computing, low latency access technologies: fiber to the home (FTTH)/fiber to the room (FTTR), Wifi6, 5G stand alone (5G SA) and beyond…) as well as extend network softwarization towards programmable networks that offer network capabilities via Network APIs making real the Network as a Service (NaaS) or Network as a Platform (NaaP) paradigm introduced in Section V. C.

Therefore, the evolution of the metaverse, web3 and blockchain will raise sustainability concerns such as their impact on carbon footprint. They could also get benefit from Network as a Platform approaches to get the most of the network to improve users´experience. These technologies will impact TEA in Comms by increasing required CapEx to deploy networks capable of coping with such increases in traffic, network performance, stability and latency, carbon price costs and OpEx, as well as requiring more complex architectures including edge computing to cope with these new technical challenges.

### 4.7 Blockchain Secured SDN

The centralized control included in software defined networking (SDN) is susceptible to Denial of Service (DoS) attacks and single failure in the control plane [18]. SDN is also subject to security challenges and therefore potential threats related to the communication and interaction of application and data planes with the control plane [18].

Latah and Kalkan propose in reference [18] a Blockchain Secured SDN (BC-Sec-SDN), a hierarchical architecture to improve SDN security and synchronization, offloading proactive decision making from SDN controllers to smart contracts on the blockchain. This novel approach brings new challenges related to the limitations of BC systems´ latency, throughput and consensus protocols. These limitations need to be overcome to meet SDN high load scenarios requirements [18].
Hence, the combination of SDN and blockchain pose new challenges for TEA in Comms, impacting on technical complexity of required architectures, as well as associated CapEx and OpEx.

## 5 Need for a Universal Techno-Economic Assessment Model?

Given this complex landscape with increasing technical complexity and exposed implications for TEA, there is a need for techno-economic assessment models that provide flexibility and adaptation capability to this dynamic context of continuous evolution and innovation, and therefore, with generalization and universal application attributes.

So the research question the author formulates is: "Is it possible to define universally applicable, scalable, flexible and generalizable techno-economic models that make it possible to compare any networking & security solutions in order to help the different market agents make agile decisions?"

In reference [3], the author studied techno-economic models for access technologies, considering the following relevant points:
- Traditionally, techno-economic models are defined as methods used to evaluate the economic viability of complex technical systems [7], ignoring an authentic evaluation of technical viability that takes into account the requirements and preferences of users, which carries the risk of committing serious technical errors that can compromise the expected economic viability.
- The techno-economic models in the literature are eminently oriented towards the dynamics of the deployment of access networks promoted by manufacturers, vendors and operators, ignoring the perspective of end users.
- Since the 1990s, different projects with public funding have been developed within the framework of different European R&D programs with the aim of developing and evolving access networks, which have given rise to most of the existing literature regarding models of techno-economic evaluation of access technologies.

The review and analysis of the literature made by the author and exposed in [3, 4], together with the wide variety of different architectural modalities and technologies, the high CapEx and OpEx required for access networks, as well as the significant volume of scientific research on techno-economics supported by EU public funding, led to conclude it was interesting to define the characteristics required to have a universal and generalizable theoretical model for techno-economic evaluation of access technologies, in order to develop a specific classification of the literature and detect areas of improvement.

After defining the characteristics and detecting areas for improvement, the author proposes in [3, 4] a new techno-economic model called Universal Techno-Economic Model (UTEM) that presents a higher degree of compliance than the models in the literature, achieving this dual main objective in his research work:

1) Defines a techno-economic model of universal, scalable, flexible and generalizable application that allows the evaluation and comparison of multiple access network technologies in different scenarios and use cases.

2) Develops a methodology for applying the techno-economic model to facilitate its use by different market agents, providing guidelines for the design of scenarios, application of the model and proper interpretation of the results obtained.

Given the complementary views exposed in previous sections and their implications for TEA, it makes sense to continue this research line to improve techno-economic assessment models to integrate and manage all this increasing complexity and contribute to agile decision-making for any market stakeholder.

## 6 Some Use Cases in Comms, Networking & Security

### 6.1 TEA models for 5G and beyond

There are numerous combinations of technologies that can compose a 5G and beyond solution: multiple radio access technologies (RATs) (e.g.: 5G New Radio, LTE, WiFi, etc.), network virtualization, new generation antennas (massive multiple input multiple output – mMIMO), multiple frequency carrier bands, cloud, edge computing, multi-access edge computing (MEC), heterogeneous deployments (HetNets) with macro-cells and small cells (micro-cells, pico-cells, femto-cells), network slicing, and so on.

There is also an increasing myriad of use cases for 5G and beyond starting from eMBB (enhanced Mobile Broadband), uLLC (ultra-Low Latency Communications), and mMTC (massive Machine-Type Communications) providing different services in different industries:

- Industry 4.0: connected factories, AGVs, Digital Twins, Cloud Robotics, Video Edge Analytics, Drones control, etc.
- TV, Media & Events: 360º video for events, automatic production, corporate events broadcasting, etc.
- eHealth: remote surgery, VR/AR aided rehabilitation, 5G music therapy, etc.
- Intelligent Mobility: intelligent automotive, platooning, V2X, etc.
- Tourism & Entertainment: AR/VR, 360º reality, holographic reality, cloud gaming, simultaneous translation, etc.
- Education: real-time learning content distribution, AR/VR/360º reality, remote real-time assistance, etc.
- Operations, Security and Emergency: Mission Critical Push-To-Talk, Mission Critical Data/Video, 5G Enhanced positioning for Mission Critical services, ProSe -Proximity Services-, remote real time-assistance, monitoring and control of operations, drones fleet management, aerial base stations deployment, etc.

5G enables AI application and digital transformation across all industries and organizations, and eventually the whole of our societies.

Many of these use cases require the deployment of 5G public networks and/or 5G private networks involving different market stakeholders (manufacturers, vendors, CSPs, telecom operators, MVNOs, cloud providers, customer enterprise and corporate organizations, retail companies, cities and town halls, regulators, and technological consultants).

Therefore, 5G different architectural flavors challenge industry stakeholders as technical complexity increases, so decision-makers need an effective and agile techno-economic assessment to choose the optimal solution.

All techno-economic assessment models for 5G in the literature were elaborated specifically for a reduced number of scenarios as reviewed in reference [10]. Besides, they lack generalization capabilities that allow their adaptation to different use cases and evolving 5G architectures. The reviewed models in [10] are not sufficiently flexible to integrate new technical and economic parameters as well as other stakeholders´ perspectives apart from telecom operators, and do not allow agile assessment through automation.

Based on a review of the literature on techno-economic assessment models for 5G, the characteristics of a theoretical 5G techno-economic assessment reference model, which considers all stakeholders' perspectives, as well as technical and economic feasibility of any 5G current and future architecture as well as automation capabilities for agile assessment, were proposed in reference [10] as 15 characteristics named *C1-C15* plus an additional one (C16) that the author includes in the present paper to consider the open source availability of any TEA model for 5G, as suggested in reference [6]. Acronyms are included for easy identification in Table 1 (*RT, NT, MB, MT, LL, IN, AN, SI, EC, DC, CR, CB, OT, MP, AU,* and *OS*):

- C1. Assessment of both RAN and Transport Networks scenarios (acronym RT).
- C2. Business viability NPV and TCO (acronym NT).
- C3. Evaluation of enhanced Mobile Broad Band (eMBB) case (acronym MB).
- C4. Evaluation of massive machine-type communications (mMTC) use case (acronym MT)
- C5. Evaluation of ultra reliable low latency communications (URLLC) use case (acronym LL).
- C6. Evaluation of Inspection KPIs (acronym IN).
- C7. Evaluation of Analytical KPIs (acronym AN).
- C8. Evaluation of Simulation KPIs (acronym SI).
- C9. The economic assessment provides OPEX, CAPEX, Revenues, and ARPU as output parameters (acronym EC).
- C10. Demand and Capacity assessment (acronym DC).

- C11. Evaluation of different degrees of centralization of RAN or CRAN scenarios (acronym CR).
- C12. Consideration of Cost-Benefit analysis for decision (acronym CB).
- C13. Overall Technical Performance (acronym OT).
- C14. Multi-perspective: including all stakeholders´ perspective (not only mobile network operator deployment perspective) (acronym MP).
- C15. Capabilities to Automate Assessment (acronym AU).
- C16. Open Source availability of TEA model (acronym OS).

Table 1 presents the classification of the reviewed models in the literature regarding their compliance with these 16 characteristics of the theoretical reference model, showing that there is ample room for improvement as the most compliant techno-economic models in the literature (see blue bars in Fig. 1) present a 50% degree of compliance with the theoretical reference model in the literature.

At the same time, Fig. 1 and Table 1 (see last column) show a 93,8% degree of compliance for the UTEM model in green, thus reducing the gap in the literature.

**Table 1.** Classification of models in the literature according to the characteristics established for the theoretical reference Techno-Economic Assessment model for 5G [10]

| Models | C1 RT | C2 NT | C3 MB | C4 MT | C5 LL | C6 IN | C7 AN | C8 SI | C9 EC | C10 DC | C11 CR | C12 CB | C13 OT | C14 MP | C15 AU | C16 OS | Total Score | Compliance |
|---|---|---|---|---|---|---|---|---|---|---|---|---|---|---|---|---|---|---|
| Theoretical Model | 1 | 1 | 1 | 1 | 1 | 1 | 1 | 1 | 1 | 1 | 1 | 1 | 1 | 1 | 1 | 1 | 16 | 100% |
| UTEM Model (Bendicho, 2016) | 1 | 1 | 1 | 1 | 1 | 1 | 1 | 1 | 1 | 1 | 1 | 1 | 1 | 1 | | | 15 | 93,8% |
| Roblot et al., 2019 | 1 | | 1 | 1 | 1 | | | 1 | 1 | 1 | 1 | | | | | | 8 | 50% |
| Oughton et al., 2019 | 1 | 1 | 1 | | | | | 1 | 1 | 1 | | | | 1 | 1 | | 8 | 50% |
| Maternia et al., 2018 | | | 1 | 1 | 1 | 1 | 1 | 1 | 1 | | | | | | | | 7 | 44% |
| mmMAGIC, 2017 | 1 | 1 | 1 | | 1 | | | | | 1 | 1 | 1 | | | | | 7 | 44% |
| J.R. Martín et al., 2019 | | | 1 | 1 | 1 | | | 1 | 1 | 1 | | | 1 | | | | 7 | 44% |
| Mesogiti et al., 2020 | 1 | | 1 | | 1 | | | 1 | 1 | 1 | | | | 1 | | | 7 | 44% |
| Smail et al., 2017 | 1 | 1 | 1 | | | | | | 1 | 1 | | 1 | | | | | 6 | 38% |
| Bouras et al., 2019 | 1 | | 1 | | | | | | 1 | 1 | 1 | 1 | | | | | 6 | 38% |
| Neokosmidis et al., 2019 | | 1 | | 1 | | | | | 1 | 1 | | 1 | | | | | 6 | 38% |
| Walia et al., 2017 | | | 1 | 1 | 1 | | | | 1 | 1 | | | | 1 | | | 6 | 38% |
| Yaghoubi et al., 2018 | 1 | 1 | 1 | | | | | | 1 | 1 | | 1 | | | | | 6 | 38% |
| Bouras et al., 2015 | 1 | | 1 | | | | | 1 | 1 | 1 | | | | | | | 5 | 31% |
| Bouras et al., 2016 | 1 | | 1 | | | | | | 1 | 1 | 1 | | | | | | 5 | 31% |
| Kolydakis & Tomkos, 2014 | 1 | | 1 | | | | | | 1 | 1 | | | | | | | 4 | 25% |
| Nikolikj & Janevski, 2014 | | | 1 | | | | | | 1 | 1 | | 1 | | | | | 4 | 25% |
| Kusuma & Suryanegara, 2019 | 1 | | 1 | | | | | | 1 | 1 | | | | | | | 4 | 25% |
| Bouras et al., 2018 | | | 1 | | | | | | 1 | 1 | | | | | | | 3 | 19% |
| Arévalo et al., 2018 | | | 1 | | | | | | 1 | | | | | 1 | | | 3 | 19% |
| | 12 | 6 | 19 | 5 | 8 | 2 | 2 | 6 | 18 | 17 | 6 | 7 | 2 | 2 | 4 | 1 | | |

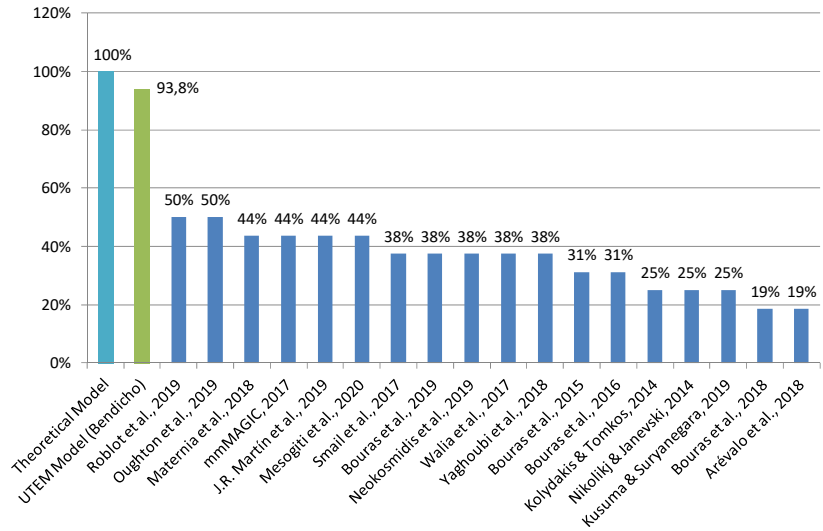

**Fig. 1.** Ranking of TEA Models for 5G considering their degree of compliance with the characteristics of the theoretical reference TEA Model for 5G [10].

As shown in the last row (heat map) of Table 1, the gap in the literature is more pronounced for C6: Evaluation of Inspection KPIs; C7: Evaluation of Analytical KPIs; C13: Overall Technical Performance; C14: Multi-perspective; C15: Capabilities to Automate Assessment (AU); and C16: Open Source Availability of the TEA model (OS).

Regarding C16, it must be stated that the UTEM model does not consider open-source availability as there is a planned technology transfer process to industry on course.

### 6.2 TEA Models for SASE, SD-WAN, SSE

The advent of software-defined networking (SDN) and network function virtualization (NFV), and more recently secure access service edge (SASE) and secure service edge (SSE), the latter driven by the extension of work-from-home (WFH) and hybrid work models derived from the SARS-Cov-2 pandemic, have created a myriad of wide-area network (WAN) vendor solutions based on any combination of these technologies.

The industry analyst firm Gartner coined the SASE concept in 2019 [11], including these 5 core networking and security functions: SD-WAN (Software Defined Wide Area Network with separated data and control planes, QoS based traffic prioritization and centralized security policies deployable with a few clicks from a central orchestrator control panel and dashboard and NFV functions deployable across the whole network on CPEs or cloud, integrating headquarters, branches, datacenters, virtual private and public clouds), Secure Web Gateway (SWG), Cloud Access

Security Broker (CASB), Zero-Trust Network Access (ZTNA), and Firewall-as-a-service (FWaaS) with the ability to identify sensitive data (DLP) or malware, decrypt content at line speed, and monitor sessions continuously for risk and trust levels.

The coexistence of legacy WAN network portions based on Multi-Protocol Label Switching (MPLS) with progressive migration to SD-WAN or SASE (full-stack vendors or multi-vendor disaggregated options) with SSE integration and the different acquisition (do-it-yourself DIY) vs. managed services subscription models (CapEx-intensive vs. OpEx-only models) from generalist providers such as a telecom operator or a managed service from a systems integrator with deep sector expertise (e.g., manufacturing, retail, financial, etc.) or regional/national expertise, increases technical complexity and multiplies the possible alternative solutions for any interested organization.

Hence, decision-makers need solvent and agile techno-economic assessment models to choose the optimal solution.

However, a review of the literature shows that there is a lack of techno-economic studies and development of TEA models oriented to this domain, except for a few papers (e.g., references [12] and [13]) that slightly cover this topic.

Considering that WAN solutions include different access network technologies for the different sites to be connected (headquarters, branches, datacenters, private and public clouds), it is possible to consider TEA models for access network technologies that leverage their extensibility capabilities to include additional technical and economic parameters related to the networking and security WAN domain (SD-WAN, SASE and/or SSE), as well as the additional parameters for the techno-economic implications of the aforementioned complementary approaches discussed in Section IV (Continuous Testing, Blockchain-based decentralized networks, Network as a Platform, and Carbon price).

Table 2 shows the ranking of different TEA models for access network technologies according to their degree of compliance with the characteristics defined for a theoretical universal TEA model [4]. Among those defined characteristics there is the "Extensibility and Flexibility" characteristic as the easiness to add new technical and economic input and output parameters.

Anyway, Table 2 shows there is a significant gap in the literature considering the compliance with the "Extensibility and Flexibility" characteristic as only UTEM model presents a degree of compliance with that characteristic higher than 0.

Therefore, there is an urgent need to bridge this gap in order to develop TEA models for SD-WAN, SASE and/or SSE domain to satisfy market needs.

**Table 2.** Ranking of TEA models for access network technologies regarding their degree of compliance with the characteristics of the theoretical reference TEA model [4]

| | Multiaccess Universality | Universality in combination of access technologies | Universality in user orientation | Universality in incorporating "micro" and "macro" approaches | Orientation to User of the Model Requirements | Geographic universality | Technical and Economic Universality | Flexibility and Extensibility | Technical and Economic Comparability | Predictive Ability | Ability to integrate with other models | Total Score | Compliance |
|---|---|---|---|---|---|---|---|---|---|---|---|---|---|
| Maximum possible score | 100 | 100 | 100 | 100 | 100 | 100 | 100 | 100 | 100 | 100 | 100 | 1100 | 100% |
| Author´s UTEM model – Bendicho (2016) | 100 | 100 | 84 | 100 | 100 | 100 | 100 | 50 | 75 | 100 | 100 | 1009 | 92% |
| Shahid & Machuca (2017) | 100 | 25 | 67 | 50 | 100 | 100 | 100 | 0 | 75 | 67 | 0 | 684 | 62% |
| Pereira & Ferreira (2009) | 100 | 25 | 67 | 50 | 50 | 100 | 75 | 0 | 50 | 100 | 0 | 617 | 56% |
| Pereira (2007) | 100 | 0 | 67 | 50 | 50 | 100 | 75 | 0 | 50 | 100 | 0 | 592 | 54% |
| Olsen et al., ECOSYS (2006) | 67 | 25 | 34 | 50 | 0 | 100 | 75 | 0 | 50 | 100 | 50 | 551 | 50% |
| Monath et al., MUSE (2005) | 67 | 25 | 34 | 50 | 0 | 100 | 75 | 0 | 50 | 100 | 50 | 551 | 50% |
| Oughton et al. (2019) | 34 | 0 | 34 | 50 | 0 | 100 | 100 | 0 | 75 | 100 | 50 | 543 | 49% |
| Feijoo et al., RURAL (2011) | 67 | 25 | 34 | 50 | 50 | 100 | 50 | 0 | 50 | 100 | 0 | 526 | 48% |
| Vergara et al., Model COSTA (2010) | 67 | 25 | 67 | 50 | 0 | 100 | 50 | 0 | 50 | 100 | 0 | 509 | 46% |
| Olsen et al., TITAN (1996) | 34 | 50 | 34 | 50 | 0 | 100 | 50 | 0 | 50 | 100 | 0 | 468 | 43% |
| Jankovich et al., EURESCOM (2000) | 67 | 25 | 34 | 50 | 0 | 100 | 50 | 0 | 50 | 100 | 0 | 476 | 43% |
| Smura, WiMAX only, TONIC & ECOSYS (2005) | 34 | 0 | 34 | 50 | 0 | 100 | 75 | 0 | 50 | 100 | 50 | 493 | 45% |
| Zagar et al. (rural broadband in Croatia) (2010) | 67 | 0 | 34 | 50 | 0 | 100 | 75 | 0 | 50 | 100 | 0 | 476 | 43% |
| Pecur, FIWI (2013) | 100 | 25 | 0 | 50 | 0 | 100 | 75 | 0 | 50 | 67 | 0 | 467 | 42% |
| Martin et al., Only HFC (2011) | 34 | 0 | 34 | 50 | 0 | 100 | 50 | 0 | 50 | 100 | 0 | 418 | 38% |
| Van der Wee et al., FTTH only, OASE (2012) | 34 | 0 | 34 | 50 | 0 | 100 | 50 | 0 | 50 | 100 | 0 | 418 | 38% |
| Van der Merwe et al., FTTH only (2009) | 34 | 0 | 34 | 50 | 0 | 100 | 75 | 0 | 50 | 67 | 0 | 410 | 37% |
| | 1106 | 350 | 726 | 900 | 350 | 1700 | 1200 | 50 | 925 | 1601 | 300 | | |

### 6.3 Cloud Cybersecurity Risk Assessment Use Case

The reader can appreciate applicability of TEA to cloud cybersecurity risk assessment (RA) in reference [14]. A review of the literature shows that cloud cybersecurity RA approaches leveraging the Cloud Security Alliance (CSA) STAR Registry that consider an organization´s security requirements present a higher degree of compliance with the defined reference model, but they still lack risk economic quantification, an aspect that can be improved by using appropriate TEA models.

## 7 TEA in request for proposals (RFP) processes

Worldwide publicly funded economic recovery instruments, such as the $1.5+ trillion in the U.S. (Build Back Better) or the €1.8+ trillion in the EU (Next Generation EU and Multiannual Financial Framework), require tremendous effort by public servants to prepare, publish and assess request for proposals (RFP).

In late June 2022, having nearly reached the first quarter of the period for 2021-2026 Recovery and Resiliency Plans (RRP) grants and loans for EU countries, less than 14% of the budget had been executed [15].

In mid October 2022, at 30,5% of the whole 2021-2026 period, only one country (Spain) had reached 44,65% of the budget execution after receiving 2[nd] payment from the EU in late July 2022. France had reached 31,8%, Greece 24,68%, Italy 23,97%, Portugal 20%, Slovakia 19,3%, while the rest (15 applicant countries) were still in the 13% budget pre-financing milestone [15].

As RFP preparation and assessment are still very handcrafted, there is an urgent need to systematize, automate and industrialize it by incorporating agile TEA tools to minimize current bottlenecks.

The authors will also conduct future research work in this direction.

## 8 Extension of TEA application to other industries

Some TEA models for Comms can also be extended to other domains and industries, such as energy, chemical, biotechnology and bioengineering optimal process design, which are key to sustainability issues to achieve the United Nations Sustainable Development Goals (SDGs).

The challenges the European Union must face towards a net-zero Europe by 2050 include reducing emissions 55 percent by 2030 [19]. This goal requires reducing emissions mainly in power, transportation, buildings, industry and agriculture by investing about €28 trillion in clean technologies and techniques over the next 30 years [19].

The huge order of magnitude of the aggregate investments required for this energy transition and the variety of possible architectural and technology pathways to reach the net-zero goal, make it necessary to apply agile and effective techno-economic assessment models in these domains and industries.

Therefore, the authors will also conduct future research work in this direction considering both the solutions deployment perspective as well as end users´ needs in order to choose the optimal solution.

## 9 Conclusions and Future work

This paper has presented a brief history of Techno-Economic Assessment (TEA) in communications (Comms), summarizing the review of the literature. It has also introduced the redefinition of TEA proposed by the author as a conclusion drawn from that previous review of the state of the art.

This article has also identified the new challenges and implications for TEA derived from the evolving technological and market context considering cloud-native virtualized networks, the Helium Network & alike blockchain-based decentralized networks, the new network as a platform (NaaP) paradigm, carbon pricing, network sharing, as well as the metaverse, web3 and blockchain technologies.

The authors have formulated the research question and shown the need to improve and develop TEA models capable to integrate and manage all the previously exposed increasing complexity.

This paper also has shown the characteristics TEA models should have and their current degree of compliance for several use cases: 5G and beyond, software-defined wide area network (SD-WAN), secure access service edge (SASE), secure service edge (SSE), and cloud cyber security risk assessment.

Considering SD-WAN, SASE and SSE use cases, the authors have identified a gap in the literature about specific TEA models for this domain, proposing the use of TEA models developed for access network technologies.

The authors have also exposed some challenges concerning request for proposal (RFP) processes and proposed TEA extensibility to that domain and other industries, including its possible key contribution to net-zero goals, given the great amount of investments required.

Therefore, there is an urgent need for agile and effective TEA in communications that allows industrialization of agile decision-making for all market stakeholders to choose the optimal solution for any technology, scenario and use case.

Moreover, techno-economic assessment can help societies face transformation with more accurate and agile decision-making tools for all stakeholders in different domains and industries, so it is key to develop research lines focused on the extensibility of the most compliant TEA models in communications to other domains and industries.

As shown in this paper, there is ample room for improvement of TEA models in communications to integrate complementary views and respond to herein exposed challenges as well as to improve performance in different use cases (5G and beyond, SD-WAN, SASE, SSE, cybersecurity risk assessment, and so on), always considering all stakeholders´ perspectives.

The authors hope their research in this direction will contribute to the general adoption of Assessment in Comms in the same way assessment in cloud adoption and migration is included in the current portfolios of service providers and consulting companies.